\documentclass[preprint,eqsecnum,aps]{revtex4}

\usepackage{graphicx}%
\usepackage{dcolumn}
\usepackage{amsmath}
\usepackage[dvips]{epsfig}      

\makeatletter
\def\btt#1{\texttt{\@backslashchar#1}}%
\DeclareRobustCommand\bblash{\btt{\@backslashchar}}%
\makeatother

\begin{document}

\title[Short Title]{ Orbital Insulators and Orbital Order-disorder 
Induced  Metal-Insulator Transition in Transition-Metal Oxides}
\author{Dong-Meng Chen$^{1,2}$ and Liang-Jian Zou$^{1}$ }
\affiliation{
\it
1 Key Laboratory of Materials Physics, Institute of Solid State Physics,
Chinese Academy of Sciences, P. O. Box 1129, Hefei 230031, China}
\affiliation{\it 2 Graduate School of the Chinese Academy of Sciences}
 
\date{\today}

\begin{abstract}
     The role of orbital ordering on metal-insulator transition in
transition-metal oxides is investigated by the cluster
self-consistent field approach in the strong correlation regime.
A clear dependence of the insulating gap of single-particle excitation
spectra on the orbital order parameter is found. The thermal 
fluctuation drives the orbital order-disorder transition, diminishes 
the gap and leads to the metal-insulator transition. The unusual 
temperature dependence of the orbital polarization in the orbital 
insulator is also manifested in the resonant x-ray scattering intensity.
\\
 

\end{abstract}


\maketitle

\newpage

\section{INTRODUCTION}

    Metal-insulator transition (MIT) in strongly correlated
transition-metal oxides (TMO) is one of the central problems in
condensed matter physics, it has attracted extensive attention in
recent decades since it has been found that high temperature
superconductivity, colossal magnetoresistance and many other phenomena 
occur in the vicinity of MIT$^{1,2)}$. These strongly correlated
electronic systems exhibit very complicated and rich phase diagrams with
temperature, doping, pressure and magnetic field $^{3,4}$. In these 
compounds, the temperature induced MIT in $V_{2}O_{3}$ and manganites 
is especially interesting, because the MIT temperature $T_{M}$ is
much smaller than the insulating gap ($\Delta$) in transport, for
example in $V_{2}O_{3}$, $T_{M}$=154 K$<<$$\Delta$ = 0.6 eV$^{5)}$.
Obviously, the thermal fluctuation is not the driven force of the
MIT. It has recently realized that the orbital degree of freedom
and orbital ordering (OO) play important roles in the groundstate 
(GS) properties of these TMO.

The OO was first proposed to explain the complicated magnetic structures 
by Kugel $et~ al$. $^{6)}$ for $KCuF_{3}$ and by Castellani 
$et~ al$. $^{7}$ for $V_{2}O_{3}$. Due to the strong
correlation between $3d$ electrons and the large anisotropy of the $3d$
wavefunctions, the orbital degree of freedom and the OO affect many
electronic and magnetic properties of these
strongly correlated systems $^{4, 6-9)}$, and have been extensively
studied in colossal magnetoresistive manganites, vanadium oxides 
and many other TMO $[$for example, see Ref.10$]$. Experimentally orbital
order-disorder transition is usually accompanied by MIT, 
such as in prototype MIT compound $V_{2}O_{3}$, an
obvious insulator to metal transition occurs at $T_{M}$$=$$T_{C}$
$^{11)}$, here $T_{C}$ is the Curie-Weiss temperature. It is believed 
that in $V_{2}O_{3}$, the OO transition (OOT), the MIT and the magnetic 
transition occur simultaneously, i.e. $T_{M} = T_{OO} = T_{C}$ $^{12)}$. 
We also notice that in lightly doped $La_{0.88}Sr_{0.12}MnO_{3}$ 
the ferromagnetic insulator to ferromagnetic metal transiting at 
T$_{M}$ is identified as an OOT by Endoh $et$ $al$ $^{13)}$, or, 
$T_{M} = T_{OO}$, ruling out the significant role of
cooperative Jahn-Teller (JT) distortion on the GS. The
orbital phase transition near the vicinity of MIT is also reported 
in $La_{1-x}Ca_{x}MnO_{3}$ ($x\approx0.2$) $^{14)}$. 
These experiments clearly establish a close relationship between OOT 
and MIT. Theoretically Castellani $et~ al$. first suggested the 
correlation between OOT and MIT in $V_{2}O_{3}$ $^{7)}$, they proposed 
that the MIT associated with the OO was driven by the variation of 
the entropy in the presence of long-range magnetic and orbital orders. 
Khaliullin $et~ al$. $^{15}$ attributed this correlation to the formation 
of orbital polaron. Nevertheless, it is not well understood theoretically 
how the OO insulator evolves to MIT with lifting temperature.

In this paper, we demonstrate that the strong orbital correlation in 
TMO with orbital degenerate 3d electrons leads the GS to be an orbital 
insulator, and the orbital order-disorder driven MIT. 
Starting from a twofold-degenerate spin-orbital interacting model, 
and using the cluster self-consistent field (cluster-SCF) approach 
developed recently, we first determine the OO GS; and then show that the
insulating gap of the single-particle spectrum opens as the long-range OO
establishes. 
With increasing temperature, the thermal fluctuation drives OOT, 
also the energy gap of the single-particle excitation vanishes at T$_{OO}$,
indicating the transition from insulator to metal. The resonant X-ray 
scattering (RXS) intensity diminishes to zero near the critical 
temperature T$_{OO}$. The rest of this paper is organized as follows: in
Sec.II we describe the effective model Hamiltonian of an orbital 
insulator and the cluster-SCF approach; then we present the magnetic
and orbital structures in GS, the T-dependent OO parameters and the gap of 
single-particle excitation spectra in Sec.III; the variation of 
the temperature and the azimuthal angle dependence of the RSX intensity
are given in Sec.IV, and the last section is devoted to the
remarks and summary.

\section{ Model Hamiltonian and Method}
\label{sec2}
  In many perovskite transition-metal oxides under the octahedral 
crystalline field, the fivefold degenerate 3d orbits of the transition 
metal ions split into lower $t_{2g}$ and higher $E_{g}$ orbits. For 
clarification and simplification we consider such an ideal cubic TMO system 
that the t$_{2g}$ orbits are filled and contribute no spin, and the twofold 
degenerate $E_{g}$ orbits are occupied by one hole or one electron,
corresponding to the electron configuration of 3d$^{9}$ in KCuF$_{3}$ 
or 3d$^{7}$ in LaNiO$_{3}$. Such a system is spin-1/2 and twofold orbital 
degenerate, or the orbital pseudospin $\tau$=1/2.
We denote the two $E_{g}$ orbits as
$|1\rangle$=$|e_{g1}\rangle$=$|d_{3z^{2}-r^{2}}\rangle$ and
$|2\rangle$=$|e_{g2}\rangle$=$|d_{x^{2}-y^{2}}\rangle$.
The major low-energy physics of the system is described by the 
twofold-degenerate Hubbard model with strong Coulomb interaction $^{6, 7)}$. 
In the second-order perturbation approximation the E$_{g}$ electrons 
interact with each other through the low-energy superexchange coupling, 
which is expressed as:
\begin{eqnarray}
    {H}_{SE}&=&\sum_{\substack{\left<ij\right>_{l}\\l=x,y,z}}
    (J_{1}\vec{s}_{i}\cdot\vec{s}_{j_{l}}
       +J_{2}I_{i}^l\vec{s}_{i}\cdot\vec{s}_{j_{l}}
       +J_{3}I_{i}^lI_{j_{l}}^l\vec{s}_{i}\cdot\vec{s}_{j_{l}}
\nonumber\\
       && ~~~~~~~~~+ J_{4}I_{i}^lI_{j_{l}}^l )
\end{eqnarray}
where the operator $I_{i}^l=\cos\left(2\pi m_{l}/3\right)\tau_{i}^z-
\sin\left(2\pi m_{l}/3\right)\tau_{i}^x$, the index l, $l= x, y$ or $z$,
denotes the direction of a bond; $\langle ij\rangle_{l}$ connects site $i$ 
and its nearest-neighbor site $j$ along the $l$ direction, and 
$(m_{x}$,$m_{y}$,$m_{z})$=$(1$,$2$,$3)$. 
$\tau^z$ and $\tau^x$ are the Pauli matrix, 
$\tau^z=\frac{1}{2}$ represents the orbital polarization in the 
state $|1\rangle$ and $\tau^z=-\frac{1}{2}$ the orbital polarization 
in the state $|2\rangle$. Thus the polarization degree of the orbit, 
$\langle\tau\rangle$, is called the orbitalization. The constants 
$J_{1},J_{2},J_{3}$ and $J_{4}$ are the superexchange interactions, and
are read:
$J_{1}=8t^2\left[U/({U}^2-J_{H}^2)-J_{H}/(U_{1}^2-J_{H}^2)\right]$,
$J_2=16t^2\left[1/(U_{1}+J_{H})+1/(U+J_{H})\right]$,
$J_{3}=32t^2\left[U_{1}/(U_{1}^2-J_{H}^2)-J_{H}/(U^2-J_{H}^2)\right]$,
and $J_{4}=8t^2\left[(U_{1}+2J_{H})/(U_{1}^2-J_{H}^2)+
J_{H}/(U^2-J_{H}^2)\right]$, with $U=U_{1}+2J_{H}$, here $4t$ is the 
hopping integral between the $|2\rangle$ orbits along the $z$ direction. 
$U$ and $U_{1}$ are the intra- and inter-orbital Coulomb interactions, 
and $J_{H}$ is the Hund's rule coupling. In this paper we adopt 
$t$= 0.1 eV and $J_{H}$= 0.9 eV.

Clearly, the fourth term in Eq.(1) is an orbital frustration interaction: 
while the exchange along the $z$ direction stabilizes the alternating 
$|3z^{2}-r^{2}\rangle$ and $|x^{2}-y^{2}\rangle$ configuration, 
the equivalent coupling along the $x$ or the $y$ directions favors other 
orbital pair configuration, i.e. $|3x^{2}-r^{2}\rangle$ and $|y^{2}-z^{2}\rangle$ 
or $|3y^{2}-r^{2}\rangle$ and $|z^{2}-x^{2}\rangle$. 
The second term in Eq.(1) is a "magnetic field" for the orbital pseudospin, 
i.e. the orbital field. Due to the relation $I_{i}^x+I_{i}^y+I_{i}^z=0$, 
the orbital field favors a peculiar orbital polarization and suppresses 
the orbital quantum frustration. However, if the spin correlations 
$\langle\vec{s}_{i}\vec{s}_{j}\rangle$ are identical along the $x, y$ and 
$z$-directions, such as in the ferromagnetic (FM) or the $\it Neel$ 
antiferromagnetic (G-type AFM) ordered structure, the orbital field vanishes.
Then the third spin-orbital interaction, which are equal along the $x$, 
the $y$ and the $z$directions, together with the fourth term 
leads to the strong quantum fluctuations and results in an orbital 
disordered ground state $^{9}$. Fortunately, the conventional JT instabiity 
lowers the cubic symmetry to the tetragonal symmetry through the JT
distortion and lifts the orbital degeneracy. The tetragonal crystalline field,
$\hat{H}_{z}=E_{z}\sum_{i}\tau_{i}^z$,
is an additional orbital field to break the symmetry of orbital space, and 
favors a peculiar orbital configuration. Thus the full Hamiltonian including 
the crystalline field term is read:
\begin{equation}
\hat{H}=\hat{H}_{SE}+\hat{H}_{z}.
\end{equation}
We will discuss the cooperative JT effect on the OO in the Sect.V.

It is still a difficult task to treat the spin and orbital correlations 
and fluctuations, and to find the GS of Eq.(2) with high accuracy. 
To study the GS properties of these TMO systems, we recently 
developed the cluster self-consistent field (cluster-SCF) approach to 
study the OO properties and the evolution of GS properties with 
interaction parameters in V$_{2}$O$_{3}$ $^{16}$. This approach combines
the exact diagonalization and the self-consistent field techniques,
the main idea is briefly addressed as follows: firstly, choose a
proper cluster, usually the unit cell of the TMO compound. Secondly, divide 
the interactions in Eq.(2) into two types: the internal bond interactions, 
H$_{ij}$, between the sites R$_{i}$ and R$_{j}$ in the cluster, and the 
external bond interactions, H$^{'}_{il}$,
between the environment site R$_{l}$ and the site R$_{i}$ in the cluster. 
The periodic condition enforces the electronic states at the environmental 
site $R_{l}$ to be identical to the corresponding atom inside the cluster. 
After diagonalizing the internal interaction H$_{ij}$ of the cluster, 
the interactions of the environment site R$_{l}$ in
H$^{'}_{il}$ acting on the cluster are obtained as a initial value of the
self-consistent field (SCF):
$h'_{i}$=$Tr_{l}(\rho_{il}H^{'}_{il})$, here $\rho_{il}$ is the
reduced density matrix of the bond $\langle il\rangle$ after
tracing over all of the other sites in the cluster.
Thirdly, diagonalize the cluster Hamiltonian H$_{ij}$ in the 
presence of the SCF $h^{'}_{i}$. Iteration is then performed until the 
orbital correlation functions and the SCF h$^{'}_{i}$ converge. 
The GS properties of the system and the effective
field of the surrounding atoms applied on the inside atoms of the
cluster are thus obtained. One of the advantages of the present approach 
is that the short-range correlation of the spins and the orbits, which is 
neglected in the conventional self-consistent mean-field method, is taken 
into account.

To check the validity and convergence of the cluster-SCF approach, we apply
it on the simple spin-1/2 AFM Heisenberg model. The Hamiltonian is read:
\begin{equation}
H=J\sum_{\left<ij\right>}\vec{s}_{i}\cdot\vec{s}_{j}
\end{equation}
where $\vec{s}_{i}$ is the operator for the $i-th$ spin and $J>0$.
In our numerical calculation code, the convergence of the GS energy 
is considered as the truncation condition of iterations, when the relative
error of the GS energy between two iterations is within $10^{-8}$, 
the GS energy is converged. The net nearest-neighbor spin correlation 
function 
$S(\vec{s}_{i},\vec{s}_{j})$$=$$\left<\vec{s}_{ i} \cdot \vec{s}_{j}
\right>-\left<\vec{s}_{i}\right>\cdot\left<\vec{s}_{j}\right>$ and  the
spin polarization $\left<s_{z}\right>$ are used cooperatively to 
determine the magnetic structure. We find 
$S(\vec{s}_{i},\vec{s}_{j})$$=$$-0.1115$ along the three axes, and the spin
polarizations of lattice clearly falls into two distinct kinds: 
$\left<s_{z}\right>$$=$$-0.4683$, and  
$\left<s_{z}\right>$$=$$0.4683$. Obviously the GS is the {\it Neel} AFM 
structure, and the averaged spin deviation of the $z$ component from 1/2 at each 
site is 0.0317, smaller than Anderson's zero-point fluctuation result 
0.078 $^{17)}$. This difference is attributed to the lack of the 
long-range quantum fluctuation of the spins in our approach, the present 
cluster-SCF approach includes only the short-range quantum fluctuations. 
This result clearly confirmed the efficiency and validity of our approach.

\section{Orbital transition induced metal-insulator transition}
\label{sec3}

In this section we first utilize the cluster-SCF method to find the magnetic
and orbital GS of Eq.(2), then explore the evolutions of the ordered orbital
parameter and the energy gap of single-particle excitation spectra with 
increasing temperature in the strong correlation limit.

\subsection{Magnetic and orbital structures in ground state}
\label{a}

   In many three-dimensional TMO the magnetic structure s are not difficult to 
determine by the neutron scattering and other experimental techniques,
in the following in accordance with the spin configuration in the experiments,
we treat the large local spins as semiclassical to save
computation time, and focus on the orbital fluctuation and the orbital 
GS for various magnetic orders. 
To determine the most stable magnetic and orbital GS structure of Eq.(2),
we choose a cubic 8-site cluster. In the cluster, each site has
three internal bonds and three external bonds. 
\begin{figure}[tbh]
\vglue 0.5cm
\epsfig{file=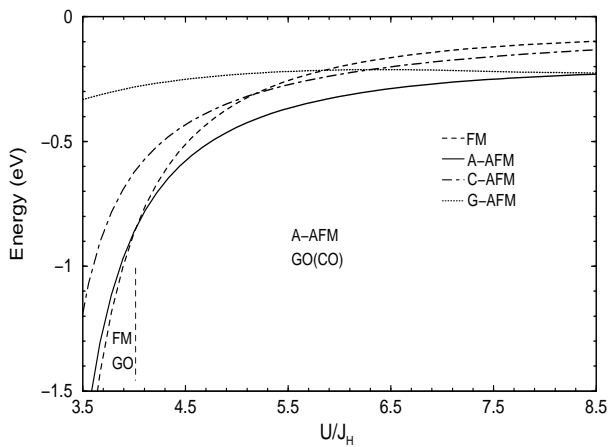,height=8.0cm,width=6.0cm,angle=270.0}
\caption{
    Dependence of the total energy on the ratio $U/J_{H}$ for different
magnetic structures. CO and GO denote the C-type and $\it Neel$ 
antiferro-orbital ordered structures, respectively. 
Theoretical parameters: $J_{H}=0.9 eV$ and $t=0.1 eV$, $E_{z}=-5 meV$.}
\label{fig:fig1}
\end{figure}
The total energy of the cluster as a function of the ratio $U/J_{H}$ for 
various magnetic structures is shown in Fig.1. It is found that for small 
$U/J_{H}$, the GS is FM, which agrees with the mean-field results of 
Roth $^{18}$ and Cyrot and Lyon-Caen's $^{19}$; while the GS 
is A-type AFM (A-AFM) for large $U/J_{H}$,
i.e. AFM coupling along the $c$-axis and FM couplings in the $ab$-plane.
In contrast, the classical approximation, which treats both the spin and the 
orbital operators as classical, shows that the GS is A-type AFM structure 
(A-AFM) over very wide $U/J_{H}$ range, as seen in Fig.2. This classical 
result is in agreement with Kugel and Khomskii's result $^{6)}$.
\begin{figure}[tp]
\vglue 0.3cm
\epsfig{file=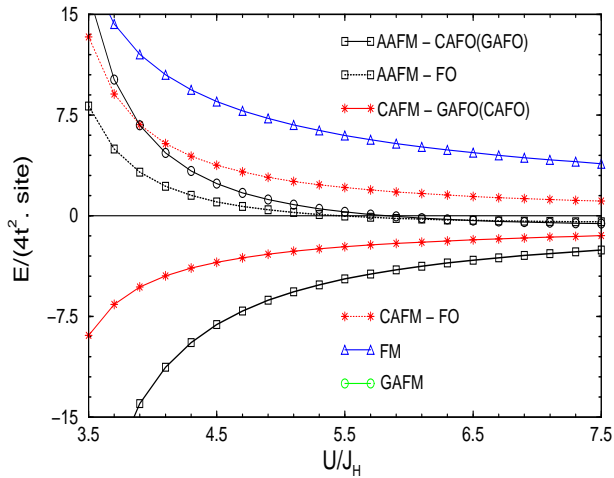,height=8.0cm,width=6.5cm,angle=270.0}
\caption{ Dependence of GS energy on the ratio $U/J_{H}$ in the classical
approximation. A-AFO denotes the A-type antiferro-orbital ordered structure.
Theoretical parameters are the same to Fig.1}
\label{fig2}
\end{figure} 
As we will show later, after taking into account the quantum effect, the 
orbital occupation and polarization are 
slightly different from the classical results.

To find the different stable magnetic structure sand the corresponding 
orbital configurations in different $U/J_{H}$ range, we take
the net orbital-orbital correlation function, 
C$_{ij}$=$\left<\vec{\tau}_{i} \cdot \vec{\tau}_{j}\right>-
\left<\vec{\tau}_{i}\right>\cdot\left<\vec{\tau}_{j}\right>$,  
together with the orbitalization $\left<\vec{\tau}\right>$ to 
determine the orbital GS. In the FM phase, the 
Ising-like orbital-orbital interactions, $I_{i}^lI_{j_{l}}^l$, which are 
identical for $l=x, y$ and $z$, as we analyzed in Sec.II, 
favors the orbital liquid GS. The tetragonal crystalline field
suppresses the quantum fluctuations, drives the electrons 
into the $|3z^{2}-r^{2}\rangle$ orbit in $E_{z}<0$, and stabilizes the GS as  
G-type antiferro-orbital (AFO) ordering, which is most favorable of the 
the anisotropic distribution of the electron clouds. This 
agrees with the empirical Goodenough-Kanamori rules $^{20)}$.
Our numerical results further show that all of the orbital correlations,
$S_{ij}$ along the x, y and z-directions, are AFO, thus each
site is AFO polarized with respect to its nearest-neighbor sites.
One finds that the orbital occupations in the two sublattices are the
combinations of $|1\rangle$ and $|2\rangle$:
$\left(\sqrt{21}|1\rangle+\sqrt{19}|2\rangle\right)/\sqrt{40}$,
and $\left(\sqrt{21}|1\rangle-\sqrt{19}|2\rangle\right)/\sqrt{40}$,
for $U/J_{H}$=4.0 and $E_{z}=-5 meV$.

With the increasing of $U/J_{H}$, the electronic superexchange interaction 
becomes small, in comparison with the crystalline field splitting, 
and more and more electrons occupy the $|x^{2}-y^{2}\rangle$ orbit. 
The FM phase becomes unstable and transits to the AFM GS at
$(U/J_{H})_{c}$=4.1.
At $U/J_{H}$=6.0, the orbital occupations in the two sublattices 
are approximately $\left(\sqrt{3}|1\rangle+|2\rangle\right)/2$
and $\left(\sqrt{3}|1\rangle-|2\rangle\right)/2$, as shown 
in Fig.3. 
\begin{figure}[tbh]
\vglue 0.5cm
\epsfig{file=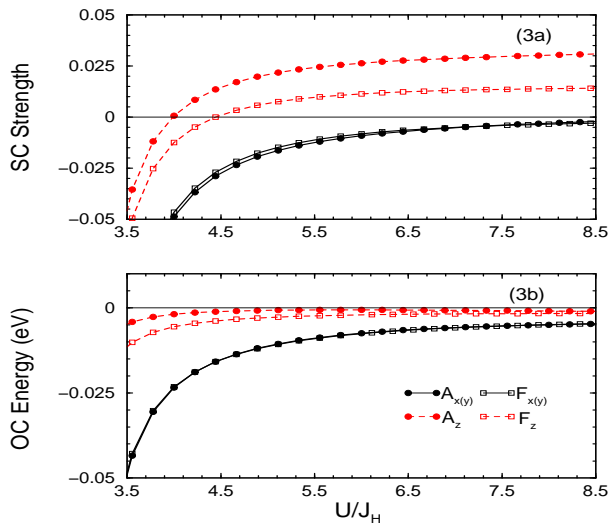,height=8.0cm,width=7.0cm,angle=270.0}
\caption{Single-bond spin coupling (SC) strength (3a) and
orbital coupling (OC) energy (3b) as the function of $U/J_{H}$
for FM and A-AFM structures. F$_{x(y,z)}$ and A$_{x(y,z)}$ denote the 
$x \left(y, z\right)$ bonds in the FM and the A-AFM structures,
respectively.}
\label{fig:fig3}
\end{figure}
The single-bond coupling energy, which is calculated for specific spin
configuration via $E^{s}_{ij}$=$Tr_{s}(\rho_{ij}H_{ij})$,
consists of the orbital-dependent spin coupling energy and the pure 
orbital coupling energy. It is found in Fig.3a and Fig.3b that for 
small $U/J_{H}$, the FM structure is the most stable,
and the quantum fluctuation of spins is small;
On the contrary, the A-AFM structure is the most stable for large $U/J_{H}$,
addressing the GS orbital configuration in typical OO compound 
$KCuF_{3}$. In the following we focus the small U/J$_{H}$ case.
More detail results for large $U/J_{H}$ and the effect of
crystalline field splitting on OO compound $KCuF_{3}$ will be presented
in a further paper.

The orbital phase transitions with the variation of the Coulomb interaction
are typical quantum phase transitions. We find that the concurrence of two 
nearest-neighbor pseudospins critically changes at the quantum transition 
point, detail result about how the entanglement of the orbital states evolves
with the quantum phase transition will be published in the future.
On the other hand, if we freeze the orbital configuration as
the G-AFO ordered phase and to search the most stable spin
configuration, we find that the magnetic structure transition
from FM to A-AFM also occurs at the same critical value
$(U/J_{H})_{c}$=4.1, confirming the validity of the orbital GS.

\subsection{Temperature dependence of orbital order parameter}
\label{b}

    After determining the orbital GS, we explore the influence of the 
thermal fluctuations on the spin and orbital order parameters. At finite 
temperature T, thermal fluctuations exciting spin waves and orbital waves 
weaken both the orbital and the spin orderings, and destroy the magnetic
order at the Curie temperature T$_{C}$ and the OO at the 
orbital critical temperature T$_{OO}$.
Since the spin order interplays with the OO, the reduction of
spontaneous magnetization with increasing temperature also softens
the orbital interaction, and {\it vice versa}. In the following we 
present the temperature dependence of the magnetic order and the OO
parameters in small $U/J_{H}$ range ($U/J_{H}$$<4.1$) in the 
conventional mean-field approximation. In the FM and G-AFO phase,
the mean-field Hamiltonian is approximated as:
\begin{equation}
   H_{MF}=\sum_{i}\left(J_{A(B)}^z\tau_{iA(B)}^z+J_{A(B)}^x\tau_{iA(B)}^x
+J_{s}s_{i}^z\right)
\end{equation}
where the coefficients $J_{A(B)}^z$, $J_{A(B)^x}$ and $J_{s}$ depend on the
magnetic order and the OO parameters, for example:
\begin{equation}
 J_{s}=3J_{1}\left<s^z\right>+3/2J_{3}\left<s^z\right>\left(\left<
\tau_{A}^z\right>\left<\tau_{B}^z\right>+\left<\tau_{A}^x\right>
\left<\tau_{B}^x\right>\right)
\end{equation}
etc. In the FM and G-AFO phase, the thermal averages of 
the spontaneous magnetization and the orbital
polarization, i.e. the orbitalization, satisfy the following
self-consistent equations:
\begin{eqnarray}
\left<s_{a}^z\right>&=&Tr\{s_{a}^z\exp\left(-\beta J_{sa}s_{a}^z\right)\}
/{Z_{a}^s}\\
\left<\tau_{A}^{x\left(z\right)}\right>&=&Tr\{\tau_{A}^{x\left(z\right)}
\exp\left(-\beta J_{A}^{x\left(z\right)}\tau_{A}^{x\left(z\right)}\right)\}
/{Z_{A}^{x\left(z\right)}}
\end{eqnarray}
where $\beta$=$1/{k_{B}T}$. Also, the self-consistent equations 
for the A-AFM and G-AFO phase at large $U/J_{H}$ can be obtained, 
but are more complicated. The temperature dependence of the spin and OO
parameters for the system with small $U/J_{H}$ is shown in Fig.4.

At low temperature, the magnetization and the two components of
the orbitalization are nearly saturated. At lifting temperature 
T$>$$T_{C}/2$, the magnetization decreases
considerably, while the orbitalization $\langle\bf\vec{\tau}\rangle$ 
almost does not change. With the temperature approaching T$_{C}$, the 
magnetization rapidly falls to zero, and the magnetic transition is 
obviously the first order, which originates from 
the strong anisotropy of the spin-orbit coupling, in agreement with T. M.
Rice's argument based the Landau phenomenal theory for multiple parameter 
orders $^{22)}$ and our renormalization group analysis to the critical
indexes of the spin-orbital interaction models $^{23)}$. Meanwhile
the OO parameter steeply diminishes a fraction at T$_{C}$, as observed 
in Fig.4. With further increasing in temperature, the OO parameter gradually 
decreases and finally vanishes at T$_{OO}$, with the characters of the 
second-order phase transition.

   Due to the existence of orbital-orbital interaction in Hamiltonian (1),
the long-range orbital correlation still exists after the magnetic
order disappears. Therefore the orbital order-disorder transition temperature 
$T_{OO}$ is higher than the magnetic transition temperature $T_{C}$, 
as observed in Fig.4. It is worthwhile noticing that the orbitalization
components $\left<\tau^{z}\right>$ and
$\left<\tau^{x}\right>$ display considerable discontinuity at the
Curie temperature $T_{C}$, which indicates the strong correlation between the
spin and the orbital degrees of freedom. The falling fraction of the 
orbitalization near T$_{C}$ is about 12$\%$ with respect to the saturated 
magnitude.
\begin{figure}[tbh]
\vglue 0.5cm
\epsfig{file=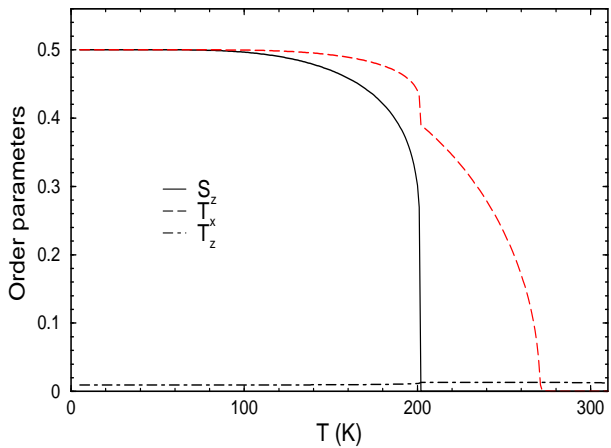, height=8.0cm, width=6.0cm, angle=270.0}
\caption{Temperature dependence of the magnetization $s_{z}$, the
orbital polarizations $\tau_{z}$ and $\tau_{x}$. Theoretical 
parameters: $U=3.6 eV$, $J_{H}=0.9 eV$, $t=0.1 eV$.}
\label{fig:fig4}
\end{figure}
The {\it ab initio} electronic structure calculations based on 
the local density approximation with correlation correction (LDA+U) 
have shown that in KCuF$_{3}$ the variation of the orbitalization from the 
low-temperature AFM phase to the high-temperature paramagnetic phase 
is small, about 5\% to 10\% $^{24)}$. Experiments, however, showed the 
variation of the orbitalization is about 50\% or even 100\% in 
KCuF$_{3}$ $^{22}$ and in V$_{2}$O$_{3}$ $^{25}$. 
This discrepancy may arise from two reasons: the first one is that the 
paramagnetic phase calculated by the LDA+U approach at zero 
temperature neglects the change of the entropy due to the variation of the
spin and orbital configurations; the second one is that in the LDA+U 
approach and the present work the long-range orbital fluctuations 
are not taken into account.

\subsection{Energy gap of single-particle spectra}
\label{c}

In this subsection we study how the gap of single-particle energy spectra of the 
E$_{g}$ electrons depends on the evolution of the OO parameter. 
In the OO phase, the long-range orbital order breaks the symmetry of the orbital 
space, and opens an insulating gap $\Delta$ in single-particle excitation 
spectra of the TMO, forming an orbital insulator. Such an insulator differs 
from the single-band Mott-Hubbard insulators in many aspects. One of
the outstanding characters is that the insulating gap strongly
depends on the OO parameter, hence closely on the temperature. 
In the TMO with quarter filling, the tight-bonding spectrum of an E$_{g}$ 
electron reads:
\begin{eqnarray}
    H_{t}&=&t\sum_{\left<i,j\right>_{x,y}}\left[d_{i1}^{\dagger}d_{j1}
\pm{\sqrt{3}\left(d_{i1}^{\dagger}d_{j2}+d_{i2}^{\dagger}d_{j1}\right)}+
3d_{i2}^{\dagger}d_{j2} \right]  \nonumber \\
   &+& 4t\sum_{\langle i,j\rangle_{z}} (d_{i1}^{\dagger}d_{j1} + h.c.)
\end{eqnarray}
where $d_{i\gamma}^{\dagger}$ creates a 3d electron at site $i$
with orbital state $\gamma$; and $\pm$ denote the signs of the
hopping integrals between two nearest-neighbor sites along the $x$ and $y$
direction, respectively. We ignore the spin index in the present FM GS.

  Strong on-site Coulomb interaction U is the main character in these 
ordered TMO, that U $>>$ t prohibits the E$_{g}$ electrons from double 
occupation at the same site, thus the single occupation constraint is 
applied on the E$_{g}$ orbits. In the limit of large Coulomb interaction, 
the constraint of no double occupancy at site $R_{i}$ is enforced by
introducing auxiliary fermions $^{26)}$, $f_{i\gamma}^{\dagger}$, and
bosons, $b_{i}$; here $f_{i\gamma}^{\dagger}$ creates a slaved
fermion (electron) at site $R_{i}$ with orbital state $\gamma$,
and $b_{i}^{\dagger}$ creates a boson (hole) at site $R_{i}$, thus
$d_{i\gamma\sigma}^{\dagger}$$=$$f_{i\gamma\sigma}^{\dagger}b_{i}$, 
and the single occupation condition is:
$\sum_{\sigma\gamma}f_{i\gamma\sigma}^{\dagger}f_{i\gamma\sigma}+
b_{i}^{\dagger}b_{i}$$=1$.
In the FM and G-AFO ordered phase, the effective low-energy Hamiltonian 
becomes:
\begin{eqnarray}
  H_{eff} &=& \sum_{k}[\epsilon_{k}^{11}f_{k1}^{\dagger}f_{k1}
+\epsilon_{k}^{22}
      f_{k2}^{\dagger}f_{k2}+\epsilon_{k}^{12}(f_{k1}^{\dagger} f_{k2}
   \nonumber\\
 &+& f_{k2}^{\dagger}f_{k1}) +\tilde{J_{1}}(f_{k+Q,1}^{\dagger}f_{k2}+
    f_{k+Q,2}^{\dagger}f_{k1}) ]  \nonumber\\
 &+& \sum_{k}\lambda (f_{k1}^{\dagger}f_{k1}+ f_{k2}^{\dagger}f_{k2}+b^{2}-1)
\end{eqnarray}
here $Q$$=$$\left(\pi,\pi,\pi\right)$, corresponding to the G-AFO order.
The average of the boson occupation is approximated as a c-number:
$<b_{i}^{\dagger}b_{i}>$$=x$. The dispersion functions are:
$ \epsilon_{k}^{11}=4xt(\cos{k_{x}}+\cos{k_{y}} $
$ +4\cos{k_{z}})+
  3\left(J_{3}\left<s^z\right>^2+J_{4}\right)(\left<\tau^z\right>+
  E_{z}/2) $,
$ \epsilon_{k}^{12}=4\sqrt{3}xt(-\cos{k_{x}} $ 
$ +\cos{k_{y}})$, and
$  \epsilon_{k}^{22}=12xt(\cos{k_{x}}+\cos{k_{y}})+
  3(J_{3}\left<s^z\right>^2+J_{4})  $
$ (\left<\tau^z\right>+E_{z}/2) $.
And the parameter 
$\tilde{J_{1}}$$=$$-3\left(J_{3}\left<s^z\right>^2+J_{4}\right)\left<\tau^x\right>$. 
In the saddle-point approximation, the constraint constant $\lambda$ 
and the boson occupation are:
\begin{eqnarray}
  \lambda &=& -\frac{1}{N}\sum_{\left<i,j\right>\gamma\gamma'}
  \left(t_{ij}^{\gamma\gamma'}\left<f_{i\gamma}^{\dagger}f_{j\gamma'}\right>+
   H\cdot{C}\right)\\
  x &=& 1-\frac{1}{N}\sum_{i}\left(\left<f_{i1}^{\dagger}f_{i1}\right>+
\left<f_{i2}^{\dagger}f_{i2}\right>\right)
\end{eqnarray}
respectively.
Physically $\lambda$ shifts the energy level of the Fermi quasiparticles,
and x gives rise to the hole concentration upon doping. In the compounds 
with quarter-filling, x $\rightarrow$ 0.
The single-quasiparticle energy spectra $E_{k}$ have four
branches and strongly depend on the OO parameter. 
For the G-AFO ordered TMO the lower two subbands are filled, since each 
magnetic and orbital unit cell contains two electrons at quarter-filling, 
the minimum of the empty subband separates from the maximum of the filled 
subbands by a gap $\Delta$. This result considerably differs from Kilian 
and Khaliullian's gapless charge excitation $^{15)}$; in fact, in their 
study the modulation of the long-range OO on the motion of the 
electrons was not taken into account.

   The thermal fluctuation weakens the OO parameter, hence softens the 
single-particle excitation spectra. The T-dependence of the energy gap of 
the single-particle spectra is shown in Fig.5.
\begin{figure}[tbh]
\vglue 0.5cm
\epsfig{file=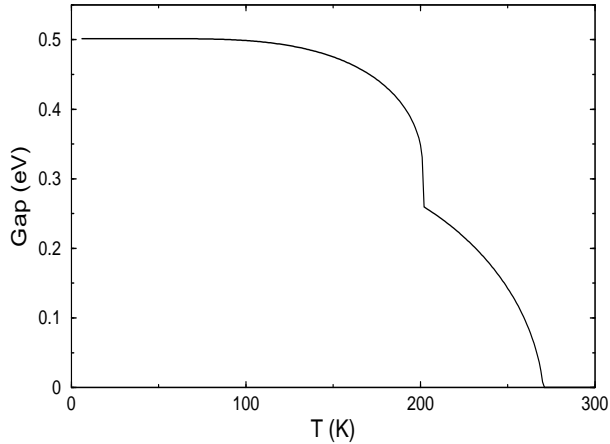,height=8.0cm,width=6.0cm,angle=270.0}
\caption{Temperature dependence of the energy gap of the single-particle 
spectra in the ferromagnetic and G-type antiferro-orbital ordered phase.
$\lambda=-0.1 eV$, and other parameters are the same to Fig.4.}
\label{fig:fig5}
\end{figure}
At low T, the strong Coulomb interaction of the 3d electrons, the large
spatial anisotropy of 3d orbits and the crystalline field splitting 
cooperatively lift the 
orbital degeneracy, and favor the ordered orbital phase. The broken
of the orbital space symmetry leads to that an electron must cost at
least $\Delta$ in energy to hop to the nearest-neighbor site.
Therefore the threshold $\Delta$ gives rise to the insulating gap
of the TMO as an orbital insulator. This energy gap manifests in the 
optical excitation and transport, and differs from the energy gap
of the orbital wave excitation.
Increasing temperature gradually destroys 
the long-range OO, and leads to the decrease of the gap of the 
separated subbands. With the further increasing of temperature to 
$T=T_{OO}$, the thermal fluctuations become so strong that the long-range 
OO and the OO parameter vanish. In this situation, the
two lower subbands merge together with the
two higher subbands, the insulating gap disappears, see Fig.5, indicating
the occurrence of MIT in TMO. Thus the energy gap of the orbital
insulator crucially depends on the OO parameter, hence the temperature.

The temperature dependence of the insulating gap $\Delta(T)$ turns out 
to be similar to that of the OO parameter, steeply diminishes a fraction 
at the Curie temperature T$_{C}$ and completely disappears at T$_{OO}$,
see Fig.4 and Fig.5.
Therefore, the temperature driven MIT in TMO is essentially induced by 
the orbital order-disorder transition. Our numerical results
demonstrate the significant discrepancy between the optical and transport
gap $\Delta(T)$ and the MIT critical temperature $T_{M}$. For the present 
system in Fig.5, the gap of the energy spectra, usually reflected in optical 
and transport experiments, is $\Delta=$ 0.5 eV; while the MIT critical 
temperature $T_{M}=T_{OO}\approx 270 K=0.023$, which accounts for the 
considerable discrepancy 
between $\Delta$ and $T_{M}$ in many orbital compounds.

\section{Temperature-dependent Resonant X-ray Scattering Intensities}
\label{sec4}

   Strong interplay between spin and orbital degrees of freedom also manifests 
In the optical excitation of orbital insulating TMO. Recently it was proposed 
to utilizing the RXS technique to measure the OO phase by
Murakami $et~ al$. $^{28}$ for manganites and Fabrizio $et~ al$. $^{29}$ for
V$_{2}$O$_{3}$. The $1s-3d$ K-edge RXS peak intensity is a useful
signal to probe the OO phase and the orbital order-disorder phase
transitions $^{11,27}$. The polarization and azimuthal
dependences of the RXS intensities provide the hidden information of the
underlying OO parameters $^{27-30)}$, and unveil the interplay between spins
and orbits. Although the signal enhancement of the
quadrupole $1s-3d$ scattering $(E_{2})$ is less than that of the
electric dipole $1s-4p$ scattering $(E_{1})$, the observed $E_{2}$
spectral line shape as a function of energy is more easy to
be identified than the quite complicated line shapes associated with
$E_{1}$ process $^{11,28,29)}$. In what follows we present 
the $E_{2}$ RXS intensity and its evolution with temperature 
based on the OO phase obtained in the preceding section.

   In the FM and G-AFO phase, the ordered 3d orbits in the sublattices 
consist of two different orbital basis:
$|\psi_{1}\rangle=\alpha_{1}|1\rangle+\alpha_{2}|2\rangle$ and $|
\psi_{2}\rangle=\alpha_{1}|1\rangle-\alpha_{2}|2\rangle$,
here the coefficients $\alpha_{1},\alpha_{2}$ are the functions of the
interaction parameters. There exist two
kinds of reflections: the fundamental reflection at
$\left(hkl\right)$ = $\left(n_{x}n_{y}n_{z}\right)$ and the
orbital superlattice reflection at $\left(hkl\right)$ =
$\left(n_{x}+\frac{1}{2}, n_{y}+\frac{1}{2}, n_{z}+\frac{1}{2}\right)$. At
the lattice symmetry-forbidden direction in which $\left(hkl\right)$ are all odd,
the orbital structure factor is written as:
\begin{eqnarray}
 F_{hkl}&=&-8\sqrt{3}f\left(\Gamma,r_{2,ds},c,\omega\right)\sqrt{n_{\epsilon k}
     n_{\epsilon'k'}}\left<\tau_{x}\right> [\epsilon_{z}k_{z}
\nonumber\\
&& \left(\epsilon_{x}^{'}k_{x}^{'}-\epsilon_{y}^{'}k_{y}^{'}\right) + 
 \left(\epsilon_{x}k_{x}-\epsilon_{y}k_{y}\right)\epsilon_{z}^{'}k_{z}^{'}]
\end{eqnarray}
where the function $f\left(\Gamma,r_{2,ds},c,\omega\right)$ is the
coefficient depending on the lifetime of the intermediate states,
$\Gamma$, the radial matrix element $r_{2,ds}$, the velocity of
photon c and the incoming photon frequency $\omega$;
$n_{\epsilon\left(\epsilon'\right) ,k\left(k'\right)}$ is the
density of the incoming (outgoing) beam of photons with
polarization $\vec{\epsilon}\left(\vec{\epsilon'}\right)$ and
wavevector $\vec{k} \left(\vec{k'}\right)$. When the incoming beam
is perfect $\sigma$-polarized, the orbital structure factor
for unrotated $\left(\sigma\sigma'\right)$ channel is:
\begin{eqnarray}
  F_{\sigma\sigma'}&=& \sqrt{\frac{n_{k\epsilon_{k}}n_{k'\epsilon_{k'}}}{3}}
     f(\Gamma,r_{2,ds},
     c,\omega) \left<\tau_{x}\right>[3\cos^2{\theta} ( \sin^2{2\varphi}
\nonumber\\
  &-&\sin{4\varphi}) +8\sin^2{\theta}(\sin^2{\varphi}\sin{2\varphi})]
\end{eqnarray}
where $\theta$ and $\varphi$ are the Bragg and the azimuthal angles, 
respectively. The azimuthal-angle and the temperature dependences of the 
RXS intensities at a selected azimuth angle of the OO superlattice Bragg 
reflection $\left(111\right)$ are shown in Fig.6.
\begin{figure}[tbh]
\vglue 0.5cm
\epsfig{file=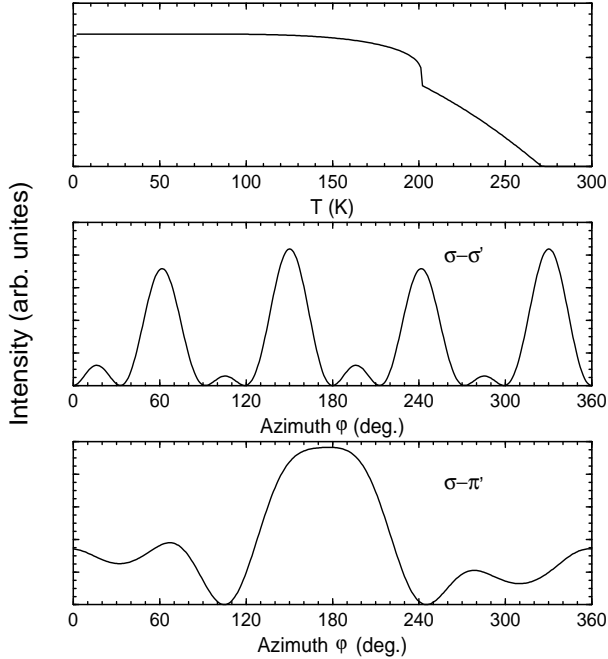,height=8.0cm,width=9.0cm,angle=270.0}
\caption{Temperature (Fig.a) and azimuthal dependence of RXS intensities in 
the ferromagnetic and G-type orbital ordered phase in (111) direction for the 
unrotated $\left(\sigma\sigma'\right)$ (Fig.b) and the rotated
$\left(\sigma\pi'\right)$ (Fig.c) channels.}
\label{fig:fig6}
\end{figure}
A steep decrease of the RXS intensities at $T_{C}$ is observed in Fig.6. 
This discontinuous decrease is associated with the lost of magnetic order, 
which diminishes a fraction of orbital interaction. Such a discontinuous 
decrease is widely observed in the RXS experiments in manganites, 
KCuF$_{3}$ and YVO$_{3}$, a significant evidence of the strong spin-orbit 
coupling. The K-edge RXS peaks disappear completely at
the OOT or MIT critical point. Therefore the RXS intensity could
be a probe to the critical point of the MIT in orbital insulators.

\section{Remarks and Summary}
\label{sec5}
   In the preceding sections we focus on the system with FM and G-AFO 
structure. For large $U/J_{H}$, the GS of the system is the A-AFM
and G-AFO (C-AFO) ordered phase, as determined by the cluster-SCF
approach. We find such an orbital ordered TMO also opens an
insulating gap in the single-particle
energy spectra, and the energy gap exhibits similar dependences on
the OO parameter and the temperature to those in the present paper. 
The dependence of the RXS intensity on temperature is also similar to the 
preceding result qualitatively, though the peak positions of the RXS
intensities slightly shift in comparison with the present
structure. Therefore different ordered orbital insulators share
many common characteristics, and are significantly different
from the conventional single-band Mott-Hubbard insulators $^{1,31)}$.

    Many cubic symmetric TMO with orbital degeneracy are inherently
unstable, the orbital degeneracy is usually lifted by lowering the crystal
symmetry through the JT distortions in LaMnO$_{3}$ $^{32}$ and 
YVO$_{3}$ $^{33}$,
or other lattice distortion modes in V$_{2}$O$_{3}$ $^{16}$. 
As far as the JT phonon-mediated orbital coupling is considered, 
an additional orbital-orbital interaction is introduced $^{9)}$:
\begin{equation}
\hat{H}_{JT}=\sum^{x,y,z}_{\substack{\left<ij_{l}\right>}}\frac{2g^{2}}{3K}
 I_{i}^lI_{j_{l}}^l
\end{equation}
where g is the electron-phonon coupling constant, and K the
restoring coefficient. For typical TMO in which JT effect
plays a role, the parameters g $\sim$ 1.2 eV, and K $\sim$ 10.0 eV. 
In this situation, the orbital correlation arises from both the
Coulomb interaction and the JT effect, thus the orbital
GS is a combination of the pure electronic spin-orbital
superexchange interaction and the crystalline field splitting 
described by Eq.(2) and the 
JT orbital interaction by Eq.(14). Utilizing the
cluster-SCF method, we find that the most stable GS magnetic
structure in this case is still FM and G-AFO ordering with a slight
modification on the orbital occupation for small $U/J_{H}$;
and for large $U/J_{H}$ the GS is A-AFM and G-AFO ordering. 
Such system also exhibits most of characters of orbital insulators.

   An obvious fact is that many OO insulators
do not exhibit MIT when the long-range OO disappears at high
temperature, such as in LaMnO$_{3}$ $^{34)}$ and YVO$_{3}$ $^{33)}$, 
except for V$_{2}$O$_{3}$. The essential reason is that these TMO are not
simple orbital insulators, there also exists strong dynamic
electron-phonon coupling, i.e., the dynamic JT effect,
even if the static cooperative JT distortion disappears at high
temperature. In these compounds, numerous dynamic JT
phonons drag the motion of 3d electrons, and localize the 3d electrons as
the incoherent polarons. The transport of the 3d electrons 
dragged by numerous dynamic phonons is of insulated polaronic character. 
Therefore these TMO are still insulators at high temperature. 
This addressed why T$_{M}$ $\neq$ T$_{OO}$ in some compounds.

  An important result from the preceding study is that the crystalline 
field splitting E$_z$ plays crucial role for the stability of the OO GS. 
A pure spin-orbital interaction can not 
solely determine the magnetic and orbital GS, since the quantum 
fluctuation is large and the spin-orbital interacting system is still 
highly degenerate. The crystalline field splitting E$_z$ may suppress
the quantum fluctuation and drive the system into an stable GS.
Similar result was also obtained by Fang and Nagaosa in the orbital compound
YVO$_{3}$ in a recent paper $^{35}$.

In summary, in an orbital-degenerate spin-orbital interacting  
system, besides the orbital wave gap, an energy gap of the electronic 
excitation is opened as the orbital order develops, and disappears with 
the vanishing of the orbital order driven by the thermal fluctuations. 
The clear dependence of the insulating gap on the orbital order parameters 
shows that such TMO possesses the orbital insulator characters, and from which 
one could interpret the considerable discrepancy between the energy gap 
in the optical and transport experiments and the MIT critical temperature.
The RXS intensity also exhibits the close interplay between spin and orbital
orders. We expect more interesting properties of orbital insulator, such as 
the critical index near the transition point of the MIT, 
will be uncovered in the further studies.

\begin{acknowledgments}
\label{acknowledgments}
One of authors Zou thanks M. Altarelli's comment. 
Supports from the NSF of China and the BaiRen project from the Chinese 
Academy of Sciences (CAS) are appreciated. Part of numerical calculation 
was performed in CCS, HFCAS.

\end{acknowledgments}

\newpage

\bibliography{apssamp}

\end{document}